          [2001/04/25 1.1 (PWD)]
\documentclass[a4paper,twocolumn]{esapub} 
\usepackage[dvips]{graphicx}

\DeclareTextSymbolDefault{\micro}{TS1}
\DeclareTextSymbol{\micro}{TS1}{181} 
\DeclareFontEncoding{TS1}{}{}
\DeclareFontSubstitution{TS1}{cmr}{m}{n}

\title{Updated results on prototype chalcogenide fibers for 10-{\micro}m wavefront spatial filtering}
\author{P. Bord\'e}
\author{G. Perrin}
\affil{LESIA, Observatoire de Paris, 5 place Jules Janssen, 92195
Meudon, France}
\author{A. Amy-Klein}
\author{C. Daussy}
\affil{LPL, Universit\'e Paris-Nord, 99 avenue Jean-Baptiste
Cl\'ement, 93430 Villetaneuse, France}
\author{G. Maz\'e}
\affil{Le Verre Fluor\'e, Campus de Ker Lann, Bruz, France}

\begin{document}


\maketitle

\begin{abstract}
The detection of terrestrial planets by Darwin/TPF missions will require extremely high quality wavefronts. Single-mode fibers have proven to be powerful beam cleaning components in the near-infrared, but are currently not available in the mid-infrared where they would be critically needed for Darwin/TPF. In this paper, we present updated measurements on the prototype chalcogenide fibers we are developing for the purpose of mid-infrared spatial filtering. We demonstrate the guiding property of our 3$^\mathrm{rd}$ generation component and we characterize its filtering performances on a 4 mm length: the far-field radiation pattern matches a Gaussian profile at the level of 3\% rms and 13\% pk-pk.
\end{abstract}

\section{Introduction}
The mid-infrared interferometer architecture for Darwin/TPF (Fridlund et al. 2000; Beichmann et al. 1999) will require extremely high-quality wavefronts. In order to relax the constraints on the quality of the optics in general, one can take advantage of spatial filtering that removes high-frequency defects in the wavefronts. This filtering can be achieved by using either pinholes or single-mode waveguides. Although single-mode waveguides are considered as much powerful devices, they are not currently available in the mid-infrared. For this reason, we have undertaken a research program to develop these components with the partnership of Le Verre Fluor\'e, a company manufacturing single-mode fluoride-glass fibers for the near-infrared.

In this paper, we present updated results with respect to the first measurements that are described by Bord\'e et al. (2002), hereafter referred to as Paper~I. In Sect.~\ref{sec:filter}, we review briefly the filtering process by single-mode fibers, then we detail the fiber's characteristics in Sect.~\ref{sec:fiber}. In Sect.~\ref{sec:setup}, we describe the test procedure and the experimental setup, before we analyze the results of our measurements in Sect.~\ref{sec:perf}.

\section{Wavefront filtering using single-mode fibers} \label{sec:filter}
Compared to pinholes that do not filter well low spatial frequency defects and have to be designed for a specific wavelength, single-mode fibers feature several advantages (see Neumann 1988 for the fundamentals). The energy coupled into the fiber is either guided into the core or radiated into the cladding. If the wavelength is above a cutoff wavelength that depends only on the fiber geometry and on materials used, the fundamental guided mode propagates alone, whereas the radiated modes get progressively eliminated. As a result, wavefront corrugations due to static aberrations or to atmospheric turbulence are filtered out.

The guiding properties of the fiber are embodied by the normalized frequency defined by
$$
V = \frac{2\pi \, a \, N\!A}{\lambda},
$$
where $a$ is the core radius, $N\!A = \sqrt{n_\mathrm{core}^2-n_\mathrm{clad}^2}$ is the numerical aperture, $n_\mathrm{core}$ and $n_\mathrm{clad}$ the refractive indices of the core and the cladding, respectively. The fiber is effectively single-mode if $V < 2.405$.

The size of the fundamental mode scales with the wavelength and thus matches nicely the size of the Airy pattern. In addition, fibers convey light and may be arranged in a beam combiner by bringing the fiber cores close to one another, as demonstrated by the FLUOR instrument in the near-infrared (Coud\'e du Foresto et al. 2002). 

\section{New prototype fiber characteristics} \label{sec:fiber}
Chalcogenides, As$_x$Se$_y$Te$_z$, are glassy materials transparent up to 12~{\micro}m that can easily be turned into fibers by drawing. A specific refractive index is obtained by adjusting the proportions $x$, $y$ and $z$. Although silver halides (AgCl, AgBr, ...) transparent up to 30~{\micro}m represent an appealing alternative choice, these polycrystals  are more difficult to turn into fiber and were not considered in the first place. Table~\ref{tab:fiber} sums up the characteristics of the sample manufactured by Le Verre Fluor\'e.
\begin{table}[htbp]
\centering
\begin{tabular}{ll}
Material           & As$_2$Se$_3$/GeSeTe$_{1.4}$ \\
Core diameter      & $2a = 23$~{\micro}m \\
Cladding diameter  & $2b = 31$~{\micro}m \\
Extra cladding     & Pb \\
Numerical aperture & NA = 0.15 \\
Cutoff wavelength  & $\lambda_c = 9.0$~{\micro}m \\
Sample length      & $l = 4$~mm \\
\end{tabular}
\caption{Prototype's characteristics.}
\label{tab:fiber}
\end{table}

A striking feature of this prototype is the fineness of the cladding together with the presence of an extra layer in lead (Pb). In the first generation sample (Paper~I), the cladding was thicker and surrounded by a protective resin layer. It turned out that the cladding-resin layer structure acted as an additional waveguide, concentric to the regular core-cladding structure, because the refractive index of the resin was smaller than that of the cladding. As a consequence, the fiber was weakly multi-mode on a 8~cm length. The solution adopted was to remove that resin layer, make the cladding diffusive (hence thinner) by chemical stripping and surround it by a mid-infrared absorbent (Pb). Hopefully, the radiated mode would get efficiently absorbed, even on a short distance. The first attempt gave promising but not quite definitive results, as the far-field radiation patterns of second generation samples still featured weak second lobes and high-frequency defects attributed to geometrical imperfections. The third generation samples studied in this paper were designed to remedy this problem.

\section{Methodology} \label{sec:setup}
\subsection{Test principle}
In addition to the spectroscopic analysis performed to test the purity of the material itself (Paper~I), the filtering property of the fiber is characterized by the measurement of the field radiated by the fiber end in the free space and at a large distance (i.e. much larger than the Rayleigh distance equal to 40~{\micro}m for our experiment). Theory predicts that the far-field radiation pattern should be the Hankel transform of the field distribution on the fiber end face. As the fundamental mode can be considered as Gaussian with an excellent approximation, the far-field radiation pattern should also exhibit a Gaussian shape (the Hankel transform of a Gaussian is a Gaussian). Moreover, the wavefronts are expected to be spherical. This technique has been applied in the past and similar experiments have been reported in the literature, e.g. Hotate \& Okoshi (1979) in the visible range.

\subsection{Instrumental setup} \label{sub:setup}
\begin{figure}[h]
\centering
\includegraphics[width=8cm]{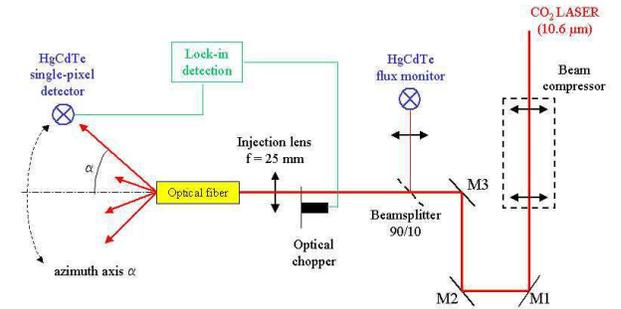}
\caption{Testbed layout.}
\label{fig:layout}
\end{figure}

We have set up a testbed at the Laboratoire de Physique des Lasers (LPL), a laboratory specialized in highly stabilized CO$_2$ lasers developed for spectroscopy and metrology purposes. A 230-Hz chopped 10.6-{\micro}m laser beam of irregular shape is focused onto the fiber head, and the radiated field at the output is measured with a single-pixel HgCdTe detector connected to a lock-in amplifier (Fig.~\ref{fig:layout}). The far-field is sampled every degree in azimuth by rotating the stage supporting the detector around a pivot straight below the fiber end. As a consequence, the detector remains on a spherical wavefront at a constant distance from the fiber end (usually 10~cm).
For our 23-{\micro}m core fiber, we expect that the energy should be radiated in cone with a half-angle subtended (width at $1/e^2$) 
$$
\theta_d = \arctan(\lambda/\pi a) \approx 16^\circ.
$$
Because the experiment was initially designed for a less divergent 40-{\micro}m core fiber ($\theta_d \approx 10^\circ$), the azimuth range is limited to $\pm \, 20^\circ$.

\section{Measurements} \label{sec:perf}
The results presented below were obtained during a 6-day measurement campaign in March 2003. Two days were necessary to rebuild the testbed, while four days were devoted to the measurements.

\subsection{Far-field radiation pattern of a pinhole}
\begin{figure}[h]
\centering
\includegraphics[width=8cm]{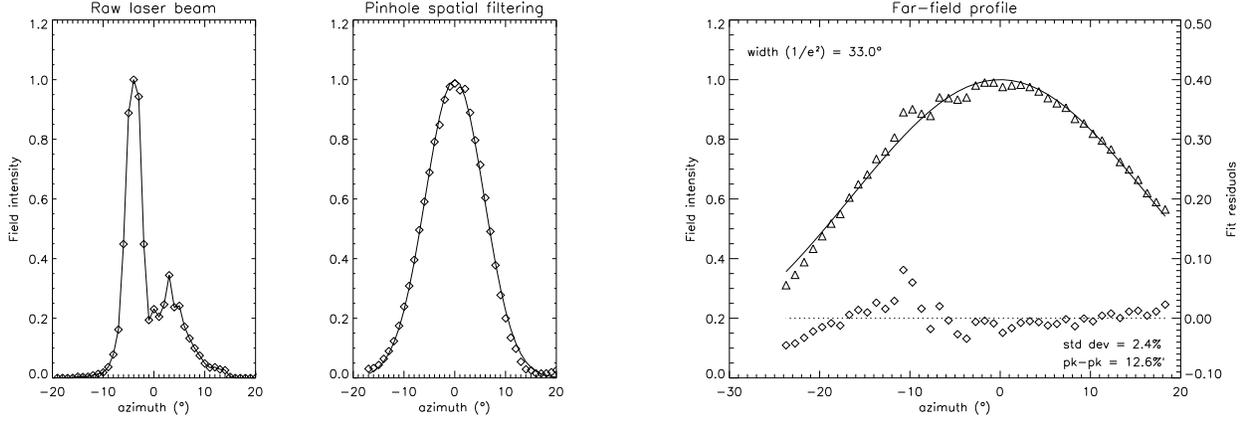}
\caption{Spatial filtering with a pinhole. Left: raw laser beam profile. Right: far-field radiation pattern obtained with a 40-{\micro}m diameter pinhole.}
\label{fig:pinhole}
\end{figure}
Before a series of measurement with the fiber, we start by recording the far-field radiation profile of a 40-{\micro}m diameter pinhole that serves as a fiducial. Figure~\ref{fig:pinhole} shows the effect of the spatial filtering on the raw laser beam.

\subsection{Filtering performances}
\begin{figure}[h]
\centering
\includegraphics[width=8cm]{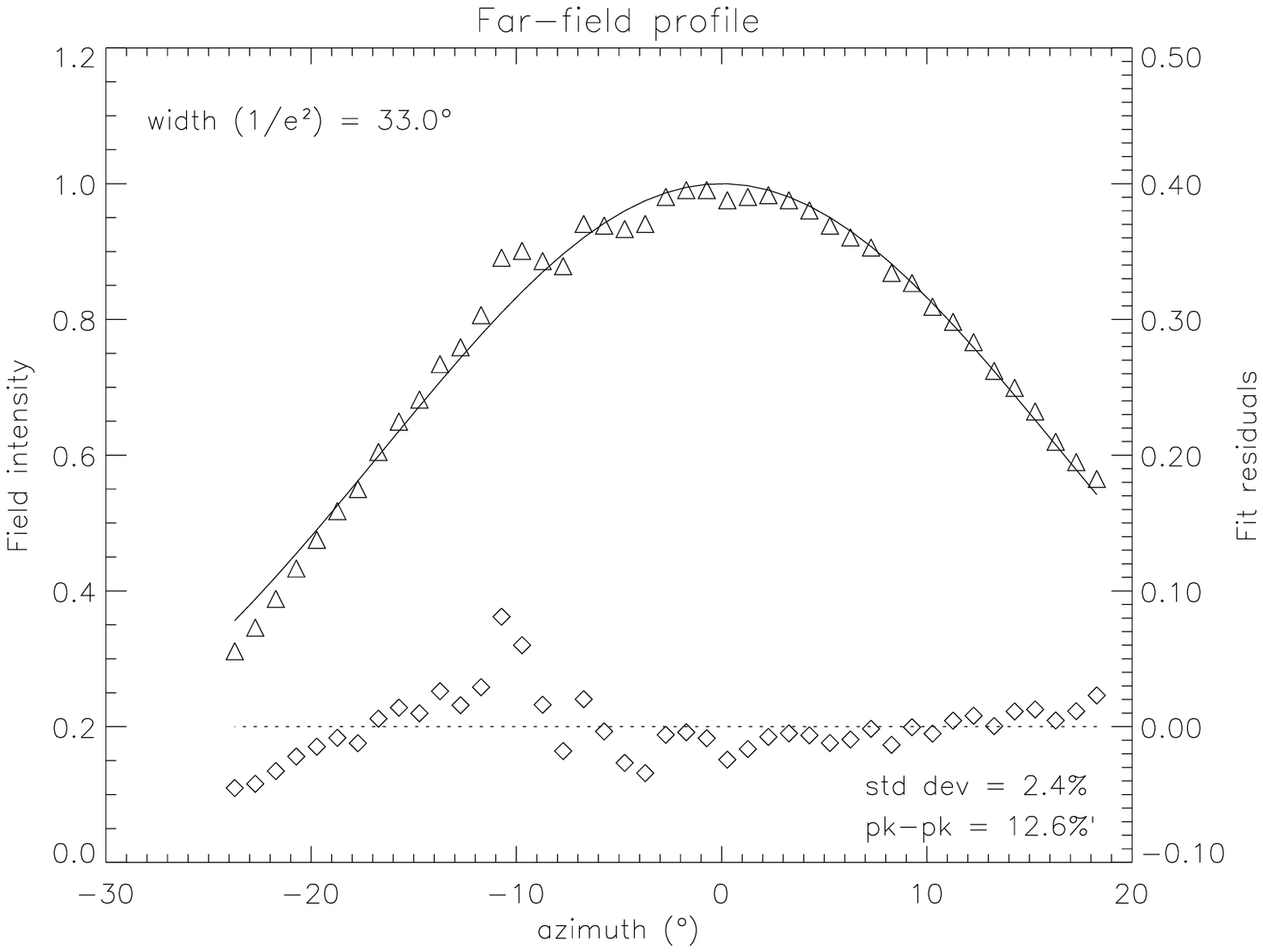}
\includegraphics[width=8cm]{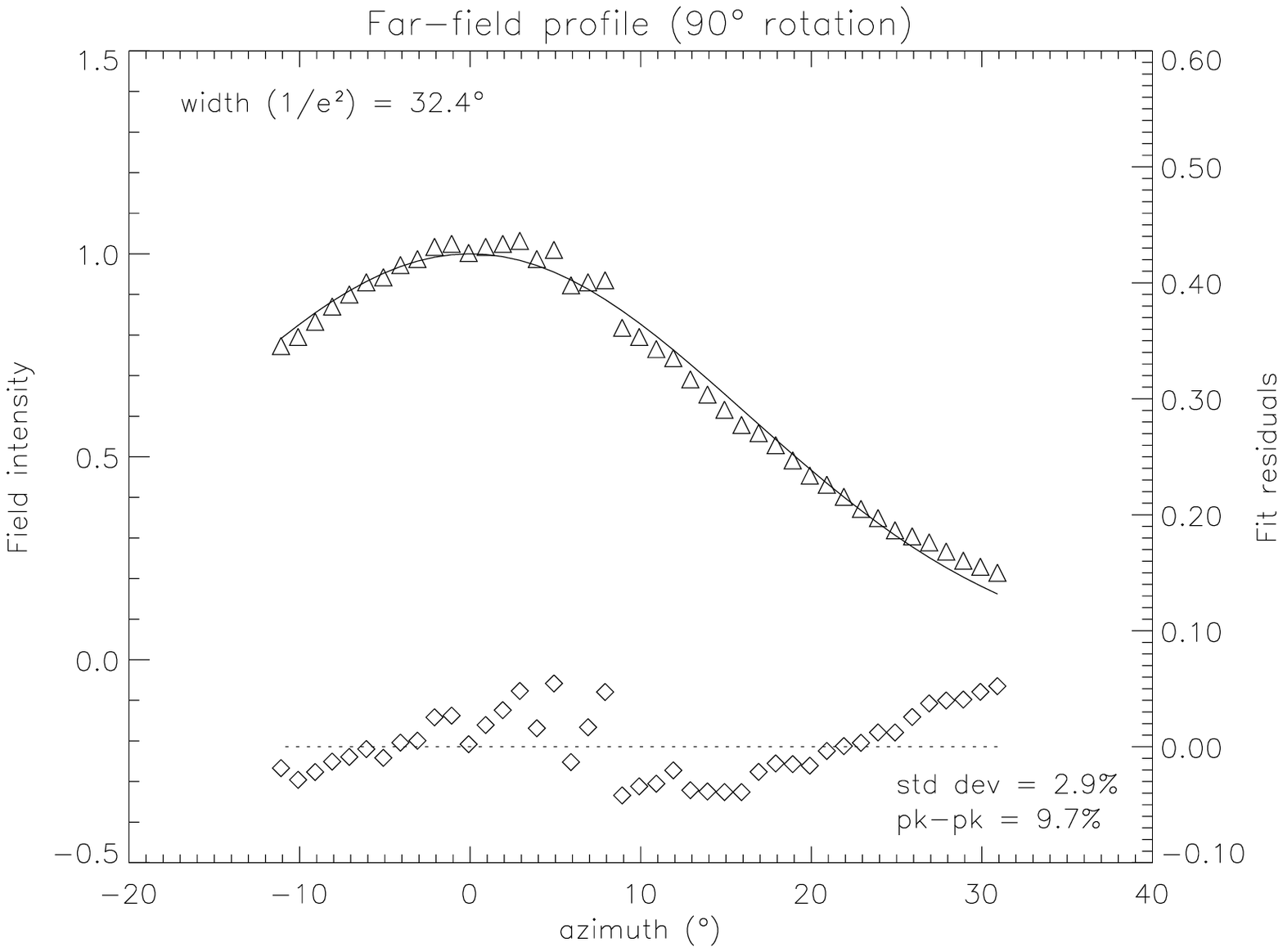}
\caption{Fiber far-field radiation profiles with best-fit Gaussians and fit residuals.}
\label{fig:fiber}
\end{figure}
Figure~\ref{fig:fiber} shows two measured profiles perpendicular to each other (the fiber was rotated by $90^\circ$ around its axis to obtain the bottom profile). These profiles show an agreement with a Gaussian to the level of 2-3\% rms and 10-13\% pk-pk (by comparison, the agreement amounts to 1.6\% rms and 5.7\% pk-pk, respectively, for our pinhole). They feature remarkably similar shapes and widths, indicating a good axial symmetry for the fiber.

Besides, if the fiber is rotated by $180^\circ$ around its axis, the measured profile happens to be identical to the one obtained for $0^\circ$: all the high-frequency defects are left unchanged. This demonstrates that these defects do not result from geometrical imperfections inside the fiber (that would have been rotation-sensitive), but rather set the limit of the spatial filtering capacity of this 4-mm sample.

\begin{figure}[h]
\centering
\includegraphics[width=8cm]{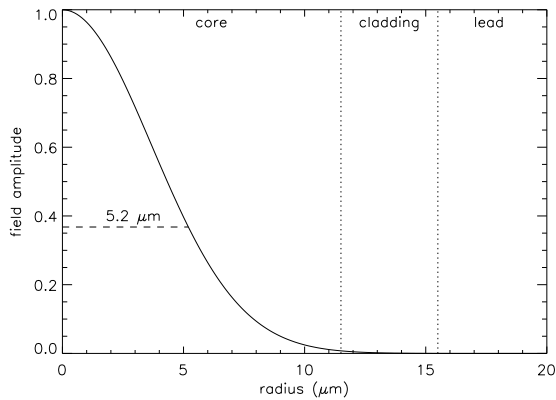}
\caption{Simple Gaussian model of the fundamental mode deduced from far-field radiation patterns displayed on Fig.~\ref{fig:fiber}.}
\label{fig:field}
\end{figure}

The profile $1/e^2$-widths, $\simeq 33^\circ$, is twice as large as expected ($\simeq 16^\circ$) for a 23-{\micro}m core fiber surrounded by an infinite cladding. In the infinite cladding case, the normalized frequency for our sample would be $V = 1.02$,
indicating that the fundamental mode would extend far into the cladding and consequently would not be well guided. However, the standard theory can not apply here since the cladding is very thin and limited by an extra layer in lead meant to absorb efficiently the mid-infrared. If we take the measured profile -- assuming that a Gaussian is a correct model (although we could not measure the wings as pointed out in Sect.~\ref{sub:setup}) -- and Hankel-transform it back, we get a 5.2-{\micro}m $1/e$-wide fundamental mode: the effect of the lead layer is to confine the field into the core (Fig.~\ref{fig:field}).

\subsection{Guiding performances}
\begin{figure}[h]
\centering
\includegraphics[width=8cm]{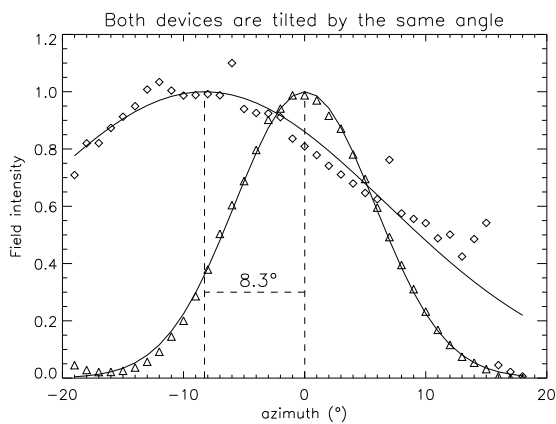}
\caption{Demonstration of the guiding property of the fiber: when a pinhole is tilted with respect to the incoming beam, the radiated profile remains centered, while it is shifted for the fiber.}
\label{fig:tilt}
\end{figure}
The stage supporting successively the pinhole and the fiber is tilted by $\simeq 10^\circ$. The far-field radiated by the fiber is shifted to the left, whereas it is not for the pinhole, thus demonstrating a guiding property for the fiber. Indeed, pinholes do not filter tip-tilt modes of the wavefront.

\subsection{Transmission}
A reliable transmission figure is usually given by the cut-back method performed by the manufacturer: the throughput of a sample is measured twice, before and after shortening one end; the difference in throughput leads to the transmission losses given the length of the missing piece. However, this method could not be applied to our already very short sample, and unfortunately our testbed was not designed to provide a good measurement of this transmission: it was not possible to disentangle the coupling efficiency from the transmission losses. It is thus only possible to assess that the fiber transmits roughly at least 30\% as much as a 31-{\micro}m pinhole.

\section{Conclusion and future work}
A new prototype single-mode fiber has been developed for the mid-infrared. It differs from classical single-mode fibers by the fact that the cladding is very thin and surrounded by a mid-infrared absorbent. By comparison with a pinhole, we have tested the guiding and the spatial filtering performances of our device. Our measurements show that (i) the fiber effectively acts as a waveguide, (ii) the effect of the outer absorbent layer is to confine the field in the fiber core, (iii) the far-field radiation pattern matches a Gaussian profile to the level of 2--3\% rms and 10--13\% pk-pk, and (iv) a longer sample (more than 4 mm) is probably needed to enhance the filtering performance. 

In the future, more studies should be done to measure the complete far-field radiation pattern (i.e. not only the central part but also the wings), determine the transmission losses and the coupling efficiency, and study the fiber's behavior with a broadband source. It should be added that this research has been conducted on a collaborative basis between Le Verre Fluor\'e and the partner laboratories since the end of the initial funding at the end of 2000. Therefore, the manufacturing of new components and any further studies presently depend on the availability of new financial supports.

\end{document}